\documentclass[onecolumn,draftcls,12pt]{IEEEtran}

\usepackage{graphicx}
\usepackage{amsmath}
\usepackage{lineno}

\begin{document}

\title{Rate Adaptation for Cognitive Radio under Interference from Primary Spectrum User}

\author{\begin{tabular}{c}
Petar Popovski$^*$, Hiroyuki Yomo$^*$, Kentaro Nishimori$^{\dagger,*}$, and Rocco Di Taranto$^*$ \\
$*$Department of Electronic Systems, Aalborg University\\
Niels Jernes Vej 12, DK-9220 Aalborg, Denmark\\
Email: \{petarp, yomo, rdt\}@es.aau.dk \\
$\dagger$ NTT Network Innovation Laboratories, NTT Corporation \\
1-1 Hikarinooka, Yokosuka-shi, 239-0847 Japan  \\
Email: \{nishimori.kentaro\}@lab.ntt.co.jp\\
\end{tabular}
\thanks{Submitted to IEEE Journal on Selected Areas in Communications
  "Cognitive Radio: Theory and Applications", March 2007}
}

\maketitle

%\pagewiselinenumbers

\begin{abstract}
A cognitive radio can operate as a secondary system in a given
spectrum. This operation should use limited power in order not to
disturb the communication by primary spectrum user. Under such
conditions, in this paper we investigate how to maximize the
spectral efficiency in the secondary system. A secondary receiver
observes a multiple access channel of two users, the secondary and
the primary transmitter, respectively. We show that, for
spectrally--efficient operation, the secondary system should apply
\emph{Opportunistic Interference Cancellation (OIC)}. With OIC,
the secondary system decodes the primary signal when such an
opportunity is created by the primary rate and the power received
from the primary system. For such an operation, we derive the
achievable data rate in the secondary system. When the primary
signal is decodable, we devise a method, based on superposition
coding, by which the secondary system can achieve the maximal
possible rate. Finally, we investigate the power allocation in the
secondary system when multiple channels are used. We show that the
optimal power allocation with OIC can be achieved through
\emph{intercepted water--filling} instead of the conventional
water--filling. The results show a significant gain for the rate
achieved through an opportunistic interference cancellation.
\end{abstract}


\section{Introduction}

A wireless network based on cognitive radio (CR)~\cite{ref:Haykin}
is allowed to \emph{reuse} the frequency spectrum which is
licensed to another system, called a primary system user. Hence,
the cognitive radio appears as a secondary user of the spectrum.
The secondary\footnote{In this text we will use the terms
``cognitive radio'' and ``secondary system'' interchangeably.}
wireless system is allowed to use certain frequency spectrum at a
certain spatial point and during a certain time, provided that it
does not cause adverse interference to the communication within
the primary system. Hence, on one hand, the operation of a
cognitive radio should be \emph{discreet} and minimally disturb
the communication in the primary system.

On the other hand, the cognitive radio should achieve a
\emph{spectrally efficient operation} and use the available
frequency in a way that is minimally disturbed by the primary
transmissions. A cognitive radio should utilize the wireless
spectrum \emph{opportunistically}~\cite{ref:Devroye_ComMag}
through frequency agility, location awareness, spectrum sensing,
rate adaptation, etc. This implies that in many cases the
cognitive radio should operate under interference from another
system and attempt to maximize its own efficiency under such a
condition. A recent work which is topically close to the
investigations presented in this paper is~\cite{ref:Devroye_IT},
where the authors analyze the information--theoretic cognitive
radio channel, defined as a 2 transmitter (TX), 2 receiver (RX)
classical information--theoretic interference
channel~\cite{ref:ThomasCover}. One of the RX--TX pairs, say the
pair 2, is a cognitive radio system, while the other system is not
necessarily a cognitive radio. The cognitive TX2 obtains \emph{a
priori} knowledge of the information that will be transmitted by
the TX1. This information is deliberately provided by TX1 and
enables TX2 to know what will be the interference when it attempts
to transmit. For such a setup, the authors derive the region of
achievable rate pairs for the two communicating pairs.

\begin{figure}
    \centering
    \includegraphics[width=6 cm]{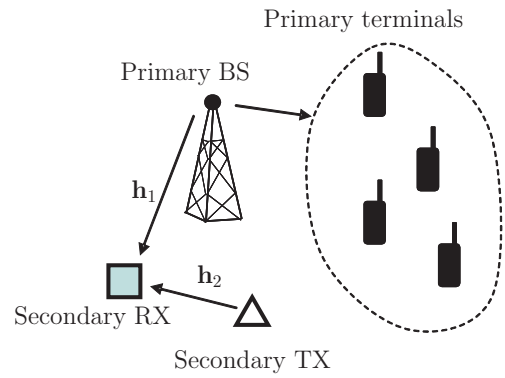}\\
    \caption{The considered scenario where the primary transmitter
    is a Primary Base Station (BS), which adapts the transmission rates to
    the population of a Primary Terminals. The Secondary
    Transmitter (TX) knows the rates used in the primary system
    and accordingly adapts its transmission to the Secondary
    Receiver (RX).
    }\label{fig:_PriSecScenario}
\end{figure}

The problem considered in this paper is essentially different from
the problem treated in ~\cite{ref:Devroye_IT}. The setup of our
problem is depicted on Fig.~\ref{fig:_PriSecScenario}. The
secondary system operates within the geographical area covered by
the primary system and using the same spectrum as the licensed
system, such that the primary and the secondary system interfere.
In order to avoid the interference towards the primary receivers,
the secondary system has a limit on the maximal transmitting
power. This limit can be decided e.~g. by using a database offered
by the primary system, where the maximal power for each particular
location is specified. Alternatively, it can be determined by
dynamic sensing of the conditions to the surrounding primary
receivers. The primary system is unaware about the existence of
the cognitive radio system and operates according to the
demands/conditions of the population of primary terminals. Thus,
we cannot assume that the cognitive transmitter has a priori
information about the messages transmitted by the primary system
and the cognitive radio should operate under the interference from
the primary system. The central question in this paper is:
\emph{Having a limited maximal power and interference from the
primary system, how to maximize the data rate in the secondary
system?} One strategy is to treat the signals from the primary
system as a noise and use only the frequency/time/space resources
where the received power from the primary is sufficiently low,
such that the secondary communication links meet the target
Signal--To--Interference--and--Noise--Ratio (SINR). Adopting such
a strategy, in our prior work~\cite{ref:Kent} we have evaluated
the spatial capacity available for communication in the CR
networks.

%\begin{figure}
%    \centering
%    \includegraphics[width=6 cm]{_PriSecScenario.eps}\\
%    \caption{The considered scenario where the primary transmitter
%    is a Primary Base Station (BS), which adapts the transmission rates to
%    the population of a Primary Terminals. The Secondary
%    Transmitter (TX) knows the rates used in the primary system
%    and accordingly adapts its transmission to the Secondary
%    Receiver (RX).
%    }\label{fig:_PriSecScenario}
%\end{figure}

The departing point in this paper is the observation that the
primary system is a legacy one, such that it is reasonable to
assume that the cognitive radio can possess the necessary system
blocks to decode the primary signals. For the scenario on
Fig.~\ref{fig:_PriSecScenario}, the secondary system attempts to
communicate during a downlink transmission from the primary Base
Station (BS). The secondary receiver (RX) receives both the signal
from the primary BS and the secondary transmitter (TX). Hence, the
secondary RX observes a \emph{multiple access channel} of two
users, one being the desired TX and the other being the undesired
primary TX. The capacity region of a multiple access channel is
defined as a region of data rates for the two users in which both
users can decode successfully. However, in the considered
scenario, the primary system adapts its data rate with respect to
the primary terminals. Such an adaptation is \emph{independent}
from the SNR at which the primary signal is received by the
secondary RX. Therefore, the secondary RX is not always able to
decode the primary signal. The cognitive system should adapt its
data rate by first considering whether the primary signal can be
decoded. This is done by observing the received powers and the
region of the achievable rates in the multiple access channel. We
call this~\emph{opportunistic interference cancellation (OIC)}, as
the decodability of the primary signal at the secondary RX depends
on the opportunity created by the (a) selection of the data rate
in the primary system and (b) the link quality between the primary
BS and the secondary RX.

In this paper we first derive the function by which the secondary
system can adapt its data rate by opportunistically cancelling the
interference from the primary system. This function is derived by
considering that the secondary system is using only one channel.
In particular, we propose a method based on superposition coding
through which any rate pair of the multiple access channel can be
achieved without time sharing, which is a method described
in~\cite{ref:ThomasCover}. This has a practical significance,
since the primary system operates independently of the secondary
system and cannot be compelled to adapt the rate in a
time--sharing manner. The rate adaptation is a function of the SNR
on the secondary link, but the parameters are the power received
from the primary and the rate applied in the primary system. It is
shown that, when the secondary system can decode the primary
system, the rate adaptation function is not a simple log--function
with respect to the power on the secondary link. In the second
part of the paper we consider a primary system that uses multiple
channels for communication and all these channels are used also by
the secondary user. With such conditions, we consider the problem
of power allocation in the secondary system in order to maximize
the sum rate achieved for all the channels. When the primary
signal on at least one of the available channels is decodable,
then the conventional water--filling cannot be used to obtain the
optimal power allocation. Instead, we introduce a method termed
\emph{intercepted water--filling} in order to obtain the maximal
sum--rate. The results confirm that there can be a significant
gain in the achievable rate when the rate adaptation is done by
opportunistic interference cancellation.

\section{System model}

We assume that each cognitive transmitter is aware about the
surrounding primary terminals and it decides the maximal power
used for transmission which guarantees that the primary receivers
will not be disturbed. The detailed discussion on the actual
methods for deciding the maximal transmitting power for the
secondary transmitters are outside of the scope of this paper. We
consider transmissions in the secondary system under the
interference from the downlink transmission in the primary system.
Analogous results can be obtained for the case of uplink
transmission in the primary system. The difference can be that,
during the uplink primary transmissions, the allowed transmission
power in the secondary system is generally higher, such that the
achievement of a spectrally efficient operation is more critical
under the interference from downlink transmissions.

Let us consider the case in which a primary system is using $M$
communication channels. A primary BS is using these channels to
transmit data to a set of primary terminals, see
Fig.~\ref{fig:_PriSecScenario}. The BS adapts the transmission
rate in each channel according to the scheduling policy and the
channel state information (CSI) of the primary terminals. We
assume that the rate adaptation in the primary system is
\emph{independent} of the activity of the secondary system.

A symbol $y_m$ received at the secondary receiver at the $m-$th
channel is given as:
\begin{equation}\label{eq:SecRXsignal}
    y_m=h_{s,m}\sqrt{{\cal E}_m}x_{s,m}+h_{p,m}x_{p,m}+z_m
\end{equation}
where:
\begin{itemize}
    \item $h_{s,m}$ is the complex channel gain on the $m-$th
    channel from the secondary TX to the secondary RX.
    \item $h_{p,m}$ is the complex channel gain on the $m-$th
    channel from the primary BS to the secondary RX.
    \item $\sqrt{{\cal E}_m}x_{s,m}$ is the signal transmitted by the secondary user on channel
    $m$, where the expected value of $x_{s,m}$ is normalized as
    $E[|x_{s,m}|^2]=1$, while ${\cal E}_m$ is proportional to the
    energy used in channel $m$.
    \item $x_{p,m}$ is the normalized signal transmitted by the primary BS channel
    $m$, such that $E[|x_{p,m}|^2]=1$
    \item $z_m$ is the complex--valued Gaussian noise with
    variance $\sigma^2$, which is identical for each channel.
\end{itemize}
We assume that the bandwidth of each channel is normalized by
setting $W=1$ [Hz], such that we can measure the time in terms of
number of symbols.

The primary system is serving the users in \emph{scheduling
epochs}. Before the starting of each epoch, the primary BS is
deciding the data rate $R_{p,m}$ which is used for transmission in
the $m-$th channel. This information is broadcasted by the BS
before the start of the scheduling epoch and is used as a preamble
for the primary user to get informed which data portion is
destined to him and what modulation/coding is used. This preamble
can be overheard by the secondary TX and RX and they can learn
about $R_{p,m}$ at each channel $m$. Let us denote by
$\beta_{p,m}$ the minimal required Signal--to--Noise Ratio for a
single link that enables successful decoding of a message sent at
rate $R_{p,m}$. Then:
\begin{equation}\label{eq:minSNR_Rpm}
    R_{p,m}=C(\beta_{p,m})
\end{equation}
where the function $C(x)$ is defined as:
\begin{equation}\label{eq:CapacityFunc}
    C(x)=\log_2(1+x) \quad \mathrm{ [bps]}
\end{equation}
Note that we should in fact use [bps/Hz], since the rate $C(x)$ is
normalized with respect to bandwidth; however, due to the
bandwidth normalization, we can use the term ``rate'' with unit
[bps] throughout the paper without causing any confusion. A
quasi--static scenario is assumed, such that a scheduling epoch
has a duration of $N$ symbols, where $N$ is sufficiently large
such that the primary BS can apply a capacity--achieving
transmission to the individual primary terminals. The primary
system is assumed to use Gaussian
codebooks~\cite{ref:ThomasCover}, which are a priori known by the
secondary system. The channel gains that $h_{s,m},h_{p,m}$ do not
change during a scheduling epoch. The secondary TX is using other
Gaussian codebooks for the secondary signal, not necessarily
related to the codebooks of the primary system.

We will use $\gamma_{s,m}$ to denote the SNR at the secondary
receiver for the signal of the secondary transmitter in the
absence of the transmission from the primary system. We will
shortly refer to it as a \emph{secondary SNR at the receiver} at
the channel $m$. Thus, we can write:
\begin{equation}\label{eq:Sec SNR at RX}
    \gamma_{s,m}=\frac{{\cal E}_m |h_{s,m}|^2}{\sigma^2}=\frac{{\cal
    E}_m}{\nu_m}
\end{equation}
where $\nu_m$ is the normalized noise energy at the $m-th$ channel
of the secondary RX. In an analogous manner, we can define the
\emph{primary SNR at the receiver} as:
\begin{equation}\label{eq:Sec SNR at RX}
    \gamma_{p,m}=\frac{|h_{p,m}|^2}{\sigma^2}
\end{equation}

From~(\ref{eq:SecRXsignal}) it follows that the transmissions of
the primary and the secondary systems are synchronized at the
secondary receiver. This enables us to consider the
information--theoretic setting of the multiple--access
channel~\cite{ref:ThomasCover}. Such a synchronization can be
achieved e.~g. through an appropriate timing advance used by the
secondary TX, without involvement of the Primary BS.

The total average energy available for secondary transmission on
all channels is:
\begin{equation}\label{eq:TotalnormEnergy}
    \sum_{m=1}^{M}{\cal E}_m={\cal E}
\end{equation}
In each
scheduling epoch, the secondary system is adapting the energy
${\cal E}_m$ and the data rate $R_{s,m}$ in each channel.

Finally, note that when we are considering a single channel
system, for simplicity we will drop the subscript $m$ from the
variables.

\section{Opportunistic Interference Cancellation (OIC)}

We will introduce the basic idea of opportunistic interference
cancellation by considering the case of a single channel, in which
the secondary transmitter allocates the total energy in each
scheduling epoch. For that purpose, we first need to consider the
achievable rates in a multiple access channel with two users.

\subsection{Two--User Multiple Access Channel}

\begin{figure}
    \centering
    \includegraphics[width=6 cm]{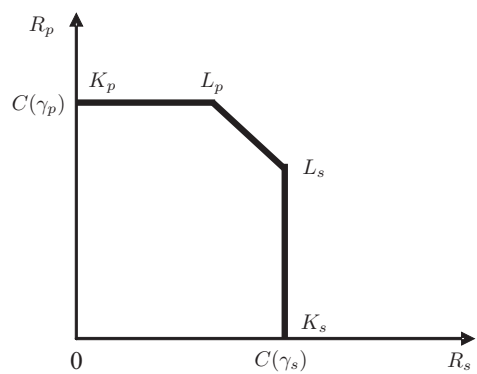}\\
    \caption{The region of achievable rate pairs ${\cal R}=(R_s,R_p)$ in a two--user multiple access channel.
    }\label{fig:_MArates}
\end{figure}

Let $\gamma_p$ and $\gamma_s$ denote the primary and the secondary
SNR at the secondary receiver, respectively. Then the secondary
receiver can reliably decode both the primary and the secondary
signal if their respective data rates $R_p$ and $R_s$ are chosen
within the convex region defined by:
\begin{eqnarray}\label{eq:MArates}
  R_s & \leq & C(\gamma_s) \\
  R_p & \leq & C(\gamma_p) \\
  R_p+R_s & \leq & C(\gamma_s+\gamma_p) \label{eq:MAratesSum}
\end{eqnarray}
This convex region is illustrated on Fig.~\ref{fig:_MArates}. The
rate pairs ${\cal R}=(R_s,R_p)$ at the points $L_s$ and $L_p$ are
given as:
\begin{eqnarray}
  {\cal R} (L_s) &=& \left( C(\gamma_s), C \left(\frac{\gamma_p}{1+\gamma_s} \right)\right) \\
  {\cal R} (L_p) &=& \left( C \left(\frac{\gamma_s}{1+\gamma_p} \right), C(\gamma_p) \right)
\end{eqnarray}
The strategies to achieve the rate pairs at the border involve
successive interference cancellation at the secondary RX. For the
rate pairs on the segment $K_sL_s$, the RX first tries to decode
the signal of the primary, treating the signal from the secondary
as an interference. After that it decodes the signal from the
secondary. An opposite strategy is used for the rates on the
segment $K_pL_p$. The suggested method in~\cite{ref:ThomasCover}
to achieve the rates on the segment $L_pL_s$ is
\emph{time--sharing}. In this case, the two transmitters should
use the rate pair ${\cal R} (L_s)$ for a fraction of time
$\theta$, and the rate pair ${\cal R} (L_p)$ for the fraction of
time $1-\theta$. By varying $\theta \in [0,1]$, any point on
$L_pL_s$ can be achieved.
%\begin{figure}
%    \centering
%    \includegraphics[width=6 cm]{_MArates.eps}\\
%    \caption{The region of achievable rate pairs ${\cal R}=(R_s,R_p)$ in a two--user multiple access channel.
%    }\label{fig:_MArates}
%\end{figure}

However, note that in the scenario that we are considering, the
rate $R_p$ of the primary is given \emph{a priori} and the
secondary TX should adapt the rate $R_s$ accordingly. As the
primary is not changing its rate during a scheduling epoch, the
usage of time--sharing is not possible and an alternative strategy
is needed to achieve the rate pairs on the segment $L_pL_s$. Let
us assume that the primary has selected the data rate to be:
\begin{equation}\label{eq:primary Rate on a Segment}
    C \left(\frac{\gamma_p}{1+\gamma_s} \right) \leq  R_p \leq C(\gamma_p)
\end{equation}
Our proposed strategy is that the cognitive transmitter should use
\emph{superposition coding}, a method used
in~\cite{ref:ThomasCover} to perform efficient broadcasting. Thus,
the secondary signal is represented as:
\begin{equation}\label{eq:Superposition Coding}
    x_s=(1-\alpha)x_{s}^{(1)}+\alpha x_{s}^{(2)}
\end{equation}
where
\begin{equation}\label{eq:Superposition Coding_2}
    0 \leq \alpha \leq 1 \qquad
    E[|x_{s}^{(1)}|^2]=E[|x_{s}^{(2)}|^2]=1
\end{equation}
such that the signal received at the secondary RX is:
\begin{equation}\label{eq:SecRXsignalSupCod}
    y=h_{s} \left ((1-\alpha)x_{s}^{(1)}+\alpha x_{s}^{(2)}\right)+h_p x_p +z
\end{equation}
The decoding of the secondary signal is performed as follows:
\begin{itemize}
    \item \textbf{Step 1:} The signal $x_{s}^{(1)}$ is decoded from $y$ by treating
    $h_{s} \alpha x_{s}^{(2)}+h_p x_p$ as an interference. After
    decoding, the signal $y^{\prime}=y-h_s(1-\alpha)x_{s}^{(1)}$
    is created.
    \item \textbf{Step 2:} The signal $x_p$ is decoded from $y^{\prime}$ by treating
    $h_{s} \alpha x_{s}^{(2)}$ as an interference. After
    decoding, the signal $y^{\prime \prime}=y^{\prime}-h_px_p$
    is created.
    \item \textbf{Step 3:} The signal $x_{s}^{(2)}$ is decoded
    from $y^{\prime \prime}$.
\end{itemize}
The coefficient $\alpha$ is determined from Step 2, by setting the
condition:
\begin{equation}\label{eq:Condition alpha}
    R_p=C \left( \frac{\gamma_p}{1+\alpha \gamma_s}\right)
\end{equation}
Recalling the definition of $\beta_p$ from (\ref{eq:minSNR_Rpm}),
we can write:
\begin{equation}\label{eq:Condition alpha beta}
    \beta_p=\frac{\gamma_p}{1+\alpha \gamma_s}
\end{equation}
such that
\begin{equation}\label{eq:Condition alpha beta final}
    \alpha=\frac{\frac{\gamma_p}{\beta_p}-1}{\gamma_s}
\end{equation}
Considering the described decoding by successive interference
cancellation, the transmission rates $R_{s}^{(1)}$ and
$R_{s}^{(2)}$ of the signals $x_{s}^{(1)}$ and $x_{s}^{(2)}$,
respectively, are chosen:
\begin{eqnarray}\label{eq:RatesSupCod}
  \nonumber R_{s}^{(1)} &=& C \left( \frac{(1-\alpha)\gamma_s}{1+\gamma_p+\alpha \gamma_s}
  \right ) \\
  R_{s}^{(2)} &=& C(\alpha \gamma_s)
\end{eqnarray}
The total rate received by the secondary user is
$R_s=R_{s}^{(1)}+R_{s}^{(2)}$. It can easily be verified that,
with rates chosen from the conditions (\ref{eq:Condition alpha})
and (\ref{eq:RatesSupCod}), the following is satisfied:
{\setlength\arraycolsep{1pt}
\begin{eqnarray}
    &R_s&+R_p = R_{s}^{(1)}+R_p+R_{s}^{(2)} = \nonumber \\
    &=& C \left( \frac{(1-\alpha)\gamma_s}{1+\gamma_p+\alpha
    \gamma_s}\right)+ C \left( \frac{\gamma_p}{1+\alpha \gamma_s}\right)+C(\alpha
    \gamma_s)=
    \nonumber \\
    &=& C(\gamma_p+\gamma_s)
\end{eqnarray}
}as required by the rate condition (\ref{eq:MAratesSum}) for the
segment $L_pL_s$. It is straightforward to prove that with the
described method we can achieve any rate point on $L_pL_s$.

\subsection{Rate Adaptation through OIC}

For the considered scenario of a multiple access channel, the
secondary TX observes the primary SNR $\gamma_p$ and the primary
data rate $R_p=C(\beta_p)$ as a priori given values. Those values
determine what is the maximal achievable rate $R_s$ when the
secondary SNR is given by $\gamma_s$. In other words, $R_s$ is a
function of $\gamma_s$ and this function is parametrized by
$\gamma_p$ and $\beta_p$:
\begin{equation}\label{eq:Rs func of gamma_s}
    R_s=F_{\gamma_p,\beta_p}(\gamma_s)
\end{equation}
For example, for $\gamma_p=0$
\begin{equation}\label{eq:Rs func gamma p 0}
    R_s=F_{\gamma_p=0,\beta_p}(\gamma_s)=C(\gamma_s)
\end{equation}
since the secondary transmitter has a non--interfered Gaussian
channel towards the receiver. The function
$F_{\gamma_p,\beta_p}(\gamma_s)$ should reflect the policy of
\emph{opportunistic interference cancellation (OIC)} for the
secondary system, where the cognitive radio makes the best
possible use of the knowledge about the interference from the
primary system. That means, if $\beta_p \leq \gamma_p$, then the
cognitive radio system can use the fact that it can decode the
primary signal in order to determine its achievable rate for given
$\gamma_s$. Alternative strategy would be the one without
Interference Cancellation (IC), where the signal from the primary
system is always treated as an undecodable interference, even when
$\beta_p \leq \gamma_p$.

In order to determine $F_{\gamma_p,\beta_p}(\gamma_s)$ we consider
two regions for $\gamma_p$:
\begin{itemize}
    \item $\gamma_p<\beta_p$. In this region the secondary
    receiver cannot decode the primary signal. Since the primary
    system is using Gaussian codebooks, the available SNR for the
    secondary signal is $\frac{\gamma_s}{1+\gamma_p}$ such that:
    \begin{equation}\label{eq:Fs gamma<beta}
        R_s=\left. F_{\gamma_p,\beta_p}(\gamma_s) \right |_{\gamma_p<\beta_p}=C\left(
\frac{\gamma_s}{1+\gamma_p} \right)
    \end{equation}
    Note that the secondary system cannot do better than this,
    since already in the achievable rate region, all the rate
    pairs on the segment $K_pL_p$ are achieved by treating the
    primary signal as an interference during the decoding of the secondary signal.
    \item $\gamma_p \geq \beta_p$. In this case the secondary
    receiver can decode the signal of the primary and use it for
    an appropriate interference cancellation. Therefore,
    the value of the $R_s$ will be chosen such that $(R_p,R_s)$
    belongs to the achievable rate region, determined for the
    given $\gamma_p$ and $\gamma_s$. In particular,
    $F_{\gamma_p,\beta_p}(\gamma_s)$ gives the maximal achievable value
    of $R_s$ for the given value of $R_p$. Depending on the value of $\gamma_s$, here we also
    differentiate two regions:
    \begin{itemize}
        \item Region $\gamma_s \leq \frac{\gamma_p}{\beta_p}-1$. In this
        region the received power from the secondary transmitter
        is such that $\gamma_s$ is low. If we plot the achievable
        rate region for the given $\gamma_p$ and $\gamma_s$, then we can
        conclude that the maximized $R_s$ lies on the line segment
        $K_sL_s$, since
        $\beta_p \leq \frac{\gamma_p}{1+\gamma_s}$. Here the receiver first decodes the
        primary signal, subtracts the decoded signal and then
        decodes the secondary signal. Hence:
        \begin{equation}\label{eq:Fs gamma_s<}
        R_s=F_{\gamma_p,\beta_p}(\gamma_s)=C(\gamma_s)
        \end{equation}
        \item Region $\gamma_s > \frac{\gamma_p}{\beta_p}-1$.
        Since in this region $\frac{\gamma_p}{1+\gamma_s} \leq \beta_p \leq
        \gamma_p$, the rate pair lies on the line segment
        $L_pL_s$. Hence, the secondary should use the transmission
        strategy based on superposition coding, described in
        the previous section. The value of $\alpha$ is determined
        according to~(\ref{eq:Condition alpha beta final}) and the
        total rate achieved by the secondary transmission can be
        written as:
        \begin{equation}\label{eq:Fs gamma_s >}
        R_s=F_{\gamma_p,\beta_p}(\gamma_s)=\log_2 \left( \frac{1+\gamma_p}{1+\beta_p}\right)+C\left(
        \frac{\gamma_s}{1+\gamma_p} \right)
        \end{equation}
    \end{itemize}
\end{itemize}

The definition of $R_s=F_{\gamma_p,\beta_p}(\gamma_s)$ can be
summarized as follows: {\setlength\arraycolsep{1pt}
\begin{equation}\label{eq:SummmaryRateFunc}
    R_s = \left\{ \begin{array}{ll}
    C\left(
    \frac{\gamma_s}{1+\gamma_p} \right) & \textrm{if $\gamma_p<\beta_p$}\\
    C(\gamma_s) & \textrm{if $\gamma_p \geq \beta_p, \gamma_s \leq \frac{\gamma_p}{\beta_p}-1$}\\
    \log_2 \left( \frac{1+\gamma_p}{1+\beta_p}\right)+C\left(
        \frac{\gamma_s}{1+\gamma_p} \right) & \textrm{if $\gamma_p \geq \beta_p, \gamma_s > \frac{\gamma_p}{\beta_p}-1$}
    \end{array} \right.
\end{equation}}

\begin{figure}
  \centering
  \includegraphics[width=9cm]{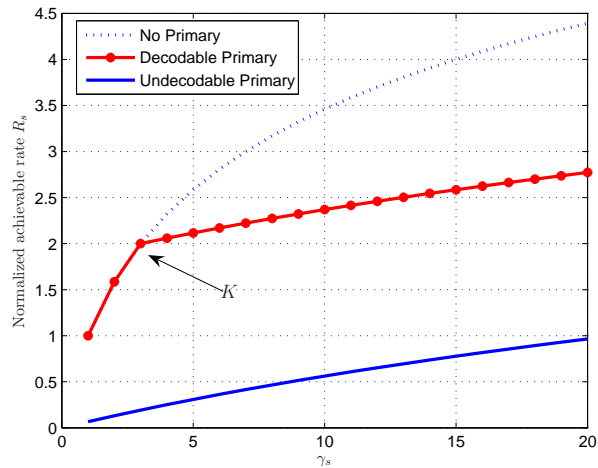}\\
  \caption{Normalized achievable rate as a function of the secondary SNR $\gamma_s$. Note that the
  abscissa and all the SNR parameters are in a linear scale. The primary SNR is
  $\gamma_p=20$, the value $\beta_p=5$ when the primary is
  decodable, while it is $\beta_p>10$ when the primary is not
  decodable.
  }\label{fig:ora}
\end{figure}

Fig.~\ref{fig:ora} exemplifies three different cases of the rate
function $F_{\gamma_p,\beta_p}(\gamma_s)$. The curve ``No
Primary'' corresponds to $\gamma_p=0$, while for the other two
curves $\gamma_p=20$. For the curve ``Decodable Primary'' the
minimal required primary SNR is $\beta_p=5$, while $\beta_p>10$
for the case ``Undecodable primary''. All mentioned SNR values are
linear, i.~e. not in [dB]. Note from the figure that, when
$\beta_p<\gamma_p$, the rate function is non--differentiable at
the point $\gamma_s=\frac{\gamma_p}{\beta_p}-1$ (the point $K$ on
the figure).
%\begin{figure}
%  % Requires \usepackage{graphicx}
%  \centering
%  \includegraphics[width=9cm]{_ora.eps}\\
%  \caption{Normalized achievable rate as a function of the secondary SNR $\gamma_s$. Note that the abscissa
%  is in the linear scale of $\gamma_s$. The primary SNR is
%  $\gamma_p=20$, the value $\beta_p=5$ when the primary is
%  decodable, while it is $\beta_p>10$ when the primary is not
%  decodable.
%  }\label{fig:ora}
%\end{figure}

\section{Extension to Multiple Channels: The Intercepted Water--Willing}

Having defined the achievable rate function
$R_s=F_{\gamma_p,\beta_p}(\gamma_s)$, we now proceed to find out
how the energy should be distributed when the secondary system has
$M>1$ communication channels.

We first consider the case $M=2$. Before stating the algorithm for
energy allocation when the rate is adapted through OIC, we first
review the conventional problem of energy/rate allocation for
parallel non--interfered Gaussian channels~\cite{ref:ThomasCover}.
If the primary signal is absent, then $\gamma_p=0$ and the problem
can be stated as follows:
\begin{eqnarray}
    \nonumber \mathrm{maximize}& C\left(\frac{{\cal E}_1}{\nu_1} \right)+C\left(\frac{{\cal E}_2}{\nu_2}
    \right)
    \\ \mathrm{for} & {\cal E}_1 \geq 0, \quad {\cal E}_2 \geq 0, \quad {\cal E}_1+{\cal
    E}_2={\cal E}
\end{eqnarray}
where $\nu_1,\nu_2$ are the normalized noise values in each
channel. Let us assume that $\nu_2>\nu_1$. By solving this
optimization problem with the Karush--Kuhn--Tucker conditions, it
can be shown~\cite{ref:ThomasCover} that this problem has the
water--filling solution, described as follows: If ${\cal E} \leq
\nu_2-\nu_1$, then ${\cal E}_1={\cal E}$ and ${\cal E}_2=0$; while
if ${\cal E} > \nu_2-\nu_1$ then ${\cal E}_1=\frac{{\cal
E}+\nu_2-\nu_1}{2}$ and ${\cal E}_2=\frac{{\cal
E}-\nu_2+\nu_1}{2}$. An interpretation of the water--filling can
be made as follows: While $C\left(\frac{{\cal E}_1}{\nu_1}
\right)$ is the faster--growing function, all the energy is poured
in channel 1; when ${\cal E}_1 = \nu_2-\nu_1$, then the rate in
both channels starts to increase with identical pace, such that
the energy $\Delta{\cal E}={\cal E}-(\nu_2-\nu_1)$ should be
equally distributed to both channels.

Let us now consider the case with the interference from the
primary and with the following values: $\nu_1=\nu_2=\nu,
\gamma_{p,1}=\gamma_{p,2}=\gamma_{p}$, while
$\beta_{p,1}=\beta_{p}>\gamma_p$, but $\beta_{p,2}<\gamma_p$. From
the discussion in the previous section, the achievable rates per
channel can be written as: {\setlength\arraycolsep{0.5pt}
\begin{eqnarray}\label{eq:Rs1Rs2waterfillingexample}
  \nonumber R_{s,1}({\cal E}_1)&=&\left\{ \begin{array}{ll}
    C\left(\frac{{\cal E}_1}{\nu}\right) \qquad \textrm{if ${\cal E}_1 \leq \nu \left(\frac{\gamma_p}{\beta_p}-1\right)={\cal E}_{10}$} \\
    \log_2 \left( \frac{1+\gamma_p}{1+\beta_p}\right)+C \left( \frac{{\cal E}_1}{\nu(1+\gamma_p)}\right)
    \quad \textrm{otherwise}
    \end{array} \right. \\
  R_{s,2}({\cal E}_2) &=& C \left( \frac{{\cal E}_2}{\nu(1+\gamma_p)}\right)
\end{eqnarray}}
The optimization problem is:
\begin{eqnarray}
    \nonumber \mathrm{maximize}& \rho_s({\cal E}_1,{\cal E}_2)=R_{s,1}({\cal E}_1)+R_{s,2}({\cal E}_2)
    \\ \mathrm{for} & {\cal E}_1 \geq 0, \quad {\cal E}_2 \geq 0, \quad {\cal E}_1+{\cal
    E}_2={\cal E}
\end{eqnarray}
However, the Karush--Kuhn--Tucker conditions cannot be directly
applied, since the function $\rho_s({\cal E}_1,{\cal E}_2)$ is not
a continuously differentiable function of $({\cal E}_1,{\cal
E}_2)$, as $R_{s,1}({\cal E}_1)$ is not a continuously
differentiable function of ${\cal E}_1$. Nevertheless, due to the
properties of the $\log$--functions, the optimal solution can be
described in the following way.

{\noindent
\begin{tabular}{p{\textwidth}}
  \hline \\
\end{tabular}
}
\noindent \emph{Region} ${\cal E} < {\cal E}_{10}$. In this
region $R_{s,1}=C\left(\frac{{\cal
    E}_1}{\nu}\right)$ and, as it grows faster than $R_{s,2}$, the
    conventional water--filling solution imposes that ${\cal E}_1={\cal
    E}$ and ${\cal E}_2=0$. For the conventional water--filling,
    such an allocation would have continued until ${\cal
    E}+\nu=\nu(1+\gamma_p)$ i.~e. ${\cal E}=\nu \gamma_p$.
    However, at ${\cal E}={\cal E}_{10}=\nu
    \left(\frac{\gamma_p}{\beta_p}-1\right)<\nu \gamma_p$ the rate
    $R_{s,1}$ starts to grow as a different function and we have
    to consider re--allocation.

\medskip

\noindent \emph{Region} ${\cal E} = {\cal E}_{10} + \Delta {\cal
E}$, where $\Delta {\cal E}>0$ is sufficiently small (we see later
what is sufficient). Let ${\cal E}_1 = {\cal E}_{10} + {\cal
E}_{11}$, such that we can write: {\setlength\arraycolsep{1pt}
\begin{eqnarray}\label{eq:Rs1insecondregion}
&R&_{s,1}=\log_2 \left( \frac{1+\gamma_p}{1+\beta_p}\right)+\log_2
\left( 1+\frac{{\cal E}_{10}+{\cal
E}_{11}}{\nu(1+\gamma_p)}\right)= \nonumber \\
&=& \log_2 \left( \frac{\gamma_p}{\beta_p}\right) +\log_2 \left(
1+\frac{{\cal E}_{11}}{\nu(1+\gamma_p)+{\cal E}_{10}}\right)
\end{eqnarray}}If we compare~(\ref{eq:Rs1Rs2waterfillingexample})
and~(\ref{eq:Rs1insecondregion}), we can conclude that $R_{s,2}$
grows with ${\cal E}_2$ faster than $R_{s,1}$ with ${\cal E}_{11}$
for all points $({\cal E}_{11},{\cal E}_{2})=(0,{\cal E}_{2})$
with $0 \leq {\cal E}_2 < {\cal E}_{10}$. Now the water--filling
solution imposes that ${\cal E}_{11}=0$ and ${\cal E}_{2}=\Delta
{\cal E}$ when $\Delta {\cal E}<{\cal E}_{10}$.

\medskip

\noindent \emph{Region} ${\cal E} = 2{\cal E}_{10} + \Delta {\cal
E}$, where $\Delta {\cal E}>0$. In this region, the energy of
${\cal E}_{10}+\frac{\Delta {\cal E}}{2}$ is allocated to each
channel.

{\noindent
\begin{tabular}{p{\textwidth}}
  \hline \\
\end{tabular}
}

\begin{figure}
  \centering
    \includegraphics[width=6 cm]{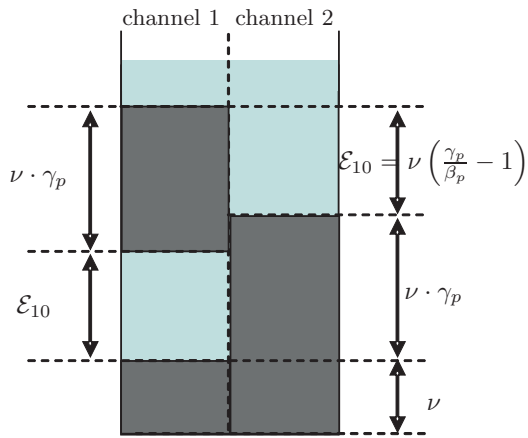}
  \caption{Example of intercepted water--filling for two channels in which $\nu_1=\nu_2=\nu$, $\gamma_{p,1}=\gamma_{p,2}=\gamma_{p}$ and
  $\beta_{p,1}=\beta_{p}>\gamma_p, \beta_{p,2}<\gamma_p$.}\label{fig:IntWFexample}
\end{figure}

The described solution is similar, yet not identical with the
water--filling solution and it can be interpreted as an
\emph{intercepted water--filling}, see
Figure~\ref{fig:IntWFexample}. Note that in the absence of the
upper ``stone'' block in channel 1, this figure would have
represented a conventional water--filling. The region pinched
between stone blocks of channel 1 and 2 can be thought of a
leakage canal of zero volume, such that while ${\cal E} < {\cal
E}_{10}$ the lower basin of channel 1 is being filled only.
%\begin{figure}
%  \centering
%    \includegraphics[width=6 cm]{_IntWFexample.eps}
%  \caption{Example of intercepted water--filling for two channels in which $\nu_1=\nu_2=\nu$, $\gamma_{p,1}=\gamma_{p,2}=\gamma_{p}$ and
%  $\beta_{p,1}=\beta_{p}>\gamma_p, \beta_{p,2}<\gamma_p$.}\label{fig:IntWFexample}
%\end{figure}

From the described interpretation of intercepted water--filling in
case of $M=2$ channels, we can devise the general solution for
power allocation when $M>2$ and the values of $\nu_m,\gamma_{p,m}$
and $\beta_{p,m}$ are arbitrary. The intercepted water--filling
produces the optimal solution. We omit the rigorous proof here and
provide only the main arguments. First, note that
$F_{\gamma_p,\beta_p}(\gamma_s)$ is always a concave function of
$\gamma_s$. When $\beta_p>\gamma_p$ the function is
non--differentiable at one point, but is still concave, as it can
be represented as a minimum of two concave
functions~\cite{ref:Rockafellar}. In that case the intercepted
water--filling implements the steepest ascent algorithm, which
leads to a globally optimal solution.

\begin{table}
    \centering
  \caption{Determining the block heights for intercepted water--filling for channels with arbitrary parameters $\nu_m,\gamma_{p,m},\beta_{p,m}$.}
  \begin{tabular}{|p{8.3cm}|}
    \hline
    \textbf{Per--channel blocks for Intercepted Water--Filling}
    \\
    \hline
    \begin{itemize}
    \item If $\gamma_{p,m}<\beta_{p,m}$, then the channel contains
    only one block of height $\nu_m(1+\gamma_{p,m})$
    \item If $\gamma_{p,m}\geq \beta_{p,m}$, then the channel
    contains two blocks. The lower block starts from the bottom
    and has a height $\nu_m$. The upper block starts at a height
    of $\nu_m+\nu_m \left(\frac{\gamma_{p,m}}{\beta_{p,m}}-1 \right)=\nu_m
    \frac{\gamma_{p,m}}{\beta_{p,m}}$. The height of the upper
    block is $\nu_m \gamma_{p,m}$.
    \end{itemize} \\
  \hline
\end{tabular}
   \label{eq:TableIWF}
\end{table}
In order to implement the intercepted water--filing, we use the
following rather visual explanation. Based on
$\nu_m,\gamma_{p,m},\beta_{p,m}$ we have to determine the height
of the ``stone'' blocks for each channel, as well as the position
of the upper stone block. This is summarized in
Table~\ref{eq:TableIWF}. Note that the upper block appears only in
the channels in which the primary signal is decodable. Having
decided the block levels/positions in the channels, the energy
allocation can be done by water--filling and considering that the
water is leaking through the side walls of the upper blocks in the
channels.

\begin{figure}
    \centering
    \includegraphics[width=6 cm]{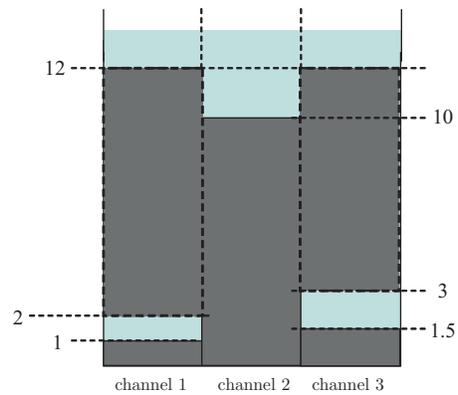}
  \caption{Example of intercepted water--filling for $M=3$ channels with
    $\nu_1=1, \nu_2=2, \nu_3=1.5$; $\gamma_{p,1}=10, \gamma_{p,2}=4,
    \gamma_{p,3}=6$; $\beta_{p,1}=10, \beta_{p,2}=4,
    \beta_{p,3}=6$. All the values are in linear scale. }\label{fig:IWFexample3ch}
\end{figure}

Rather than giving the precise algorithmic steps for intercepted
water--filling, we illustrate it by the example on
Fig.~\ref{fig:IWFexample3ch}. The chosen parameters for the
channels are $\nu_1=1, \nu_2=2, \nu_3=1.5$; $\gamma_{p,1}=10,
\gamma_{p,2}=4,\gamma_{p,3}=6$; $\beta_{p,1}=10, \beta_{p,2}=4,
\beta_{p,3}=6$. If the total energy is:
\begin{itemize}
    \item ${\cal E} \leq 0.5$: all the energy is allocated to
    channel 1.
    \item $0.5 < {\cal E} \leq 1.5$: then $\frac{{\cal
    E}+0.5}{2}$ is allocated to channel 1 and $\frac{{\cal
    E}-0.5}{2}$ is allocated to channel 3.
    \item $1.5 < {\cal E} \leq 2.5$: energy $1$ is allocated to
    channel 1 and ${\cal E}-1$ is allocated to channel 3
    \item $2.5 < {\cal E} \leq 4.5$: energy $1$ is allocated to
    channel 1, energy $1.5$ is allocated to channel 3 and energy
    ${\cal E}-2.5$ is allocated to channel 2.
    \item ${\cal E} > 4.5$: channel 1 gets $1+\frac{{\cal E}-4.5}{3}$, channel 2 gets $2+\frac{{\cal E}-4.5}{3}$,
    channel 3 gets $1.5+\frac{{\cal E}-4.5}{3}$.
\end{itemize}

Note that the total height of the blocks in a channel is equal to
$\nu_m(1+\gamma_{p,m})$. This implies that, when the amount of
energy is sufficiently high, such that the water--filling goes
above the uppermost block (in this example ${\cal E} > 4.5$), then
the power allocation of the intercepted water--filling is
identical with the allocation of the conventional water--filling.

%\begin{figure}
%    \centering
%    \includegraphics[width=6 cm]{_IWFexample3ch.eps}
%  \caption{Example of intercepted water--filling for $M=3$ channels with
%    $\nu_1=1, \nu_2=2, \nu_3=1.5$; $\gamma_{p,1}=10, \gamma_{p,2}=4,
%    \gamma_{p,3}=6$; $\beta_{p,1}=10, \beta_{p,2}=4,
%    \beta_{p,3}=6$. All the values are in linear scale. }\label{fig:IWFexample3ch}
%\end{figure}

\section{Numerical Results}

In this section we will provide a numerical illustration of the
OIC in order to show the introduced gain as compared to the case
when the primary interference is treated only as a noise.

\begin{figure}
  \centering
  \includegraphics[width=9cm]{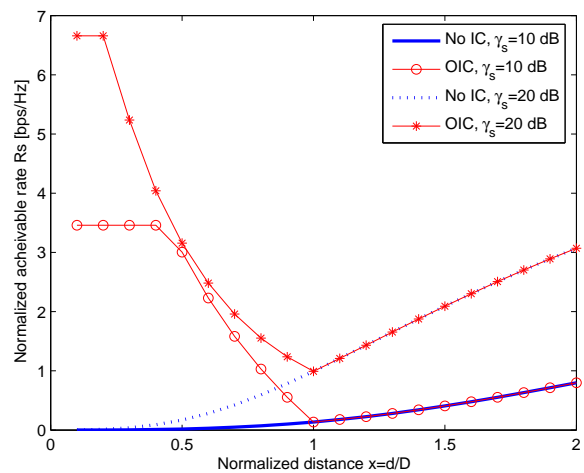}\\
  \caption{Normalized achievable rate as a function of the normalized distance of the secondary RX from the primary BS.
  The ``No IC'' case is achieved rate without Interference
  Cancellation, while OIC denotes Opportunistic Interference
  Cancellation. Here $\beta_p=20$ [dB], propagation coefficient is
  $v=3$.
  }\label{fig:LineRate}
\end{figure}

Let us first consider a scenario with $M=1$ with the following
setup. The primary system has a range of $D$ meters and it adjusts
its power so as to have a predefined SNR of $\beta_p$ for a
receiver at a distance $D$ which has a Line--of--sight (LOS) link
to the BS. Let us now consider a secondary receiver at a distance
$d$ and let $x=\frac{d}{D}$ which also has a LOS to the BS. Then
the primary SNR at the distance $d=xD$ is equal to:
\begin{equation}\label{eq:PrimarySNRLineScenario}
    \gamma_p(x)=\frac{\beta_p}{x^v}
\end{equation}
where $v$ is the propagation coefficient. Let us assume that the
secondary TX adjusts the power within the allowed range, such that
the secondary SNR at the secondary RX is $\gamma_s$.
Fig.~\ref{fig:LineRate} depicts the normalized achievable rate as
a function of the normalized distance $x$. Two values of
$\gamma_s$ are used, $10$ and $20$ dB, respectively and $\gamma_s$
is a measure of the power applied in the secondary system. For
each $\gamma_s$, two rate curves are plotted, without Interference
Cancellation (No IC) and with opportunistic Interference
Cancellation (OIC). Clearly, OIC leads to higher rate when $x<1$,
but is identical to the case without interference cancellation for
$x>1$, as the primary signal cannot be decoded when the secondary
RX is at distances $d>D$. For the OIC curves, the rate points in
the region $\frac{1}{(1+\gamma_s)^{\frac{1}{v}}} < x < 1$ are
achieved by the described strategy of superposition coding. It is
very interesting to notice that the two OIC curves are close to
each other for $x$ around $0.5$. This means that the increase of
the power for 10 dB in the secondary system has produced a small
rate increase. On the other hand, for the region $x>1$, the rate
gain out of the 10 dB improvement in the SNR is more pronounced.
Recall from~(\ref{eq:Condition alpha beta final})
and~(\ref{eq:RatesSupCod}) that the rate of the secondary signal
that is decoded after decoding of the primary signal is equal to
$C\left(\frac{\gamma_p}{\beta_p}-1 \right)$ and does not depend on
$\gamma_s$. Thus, we can conclude that for $x$ around $0.5$, this
signal carries the dominant portion of the data in the secondary
system. On the other hand, the first layer of the superposition
coding (the one decoded before the primary signal) brings rate
improvement for values of $x$ closer to the edges of the observed
region.
%Finally, for $x<0.5$, the impact of the primary
%interference is decreased, such that the increase of $\gamma_s$
%for 10 dB leads to significant rate improvement.

%\begin{figure}
%  % Requires \usepackage{graphicx}
%  \centering
%  \includegraphics[width=9cm]{LineRate.eps}\\
%  \caption{Normalized achievable rate as a function of the normalized distance of the secondary RX from the primary BS.
%  The ``No IC'' case is achieved rate without Interference
%  Cancellation, while OIC denotes Opportunistic Interference
%  Cancellation. Here $\beta_p=20$ [dB], propagation coefficient is
%  $v=3$.
%  }\label{fig:LineRate}
%\end{figure}

\begin{figure}
  \centering
  \includegraphics[width=9cm]{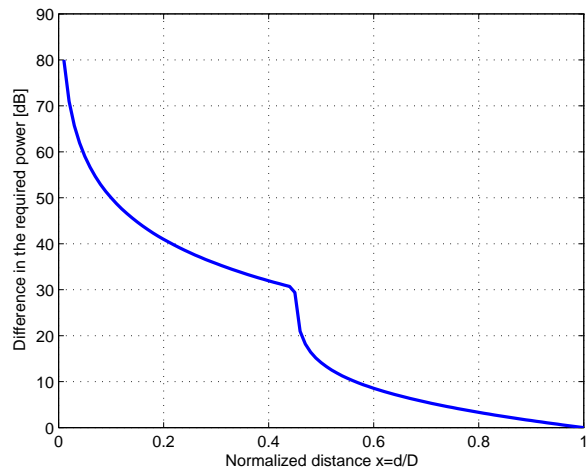}\\
  \caption{Difference in [dB] between the required power with and
  the power without OIC, respectively, in order to achieve a
  secondary rate of $C(10)$ [bps]. Here $\beta_p=20$ [dB], propagation coefficient is
  $v=3$.
  }\label{fig:LinePowerDIff}
\end{figure}

Another perspective for the same scenario is given by
Fig.~\ref{fig:LinePowerDIff}. We assume that the secondary system
aims to achieve a data rate equal to $C_0$ [bps]. Let
$\gamma_{s}^{no OIC}(C_0)$ and $\gamma_{s}^{OIC}(C_0)$ denote the
required secondary SNR to achieve $C_0$ without and with OIC,
respectively. The figure plots $\left( \gamma_{s}^{no
OIC}(C_0)-\gamma_{s}^{OIC}(C_0)\right)$ [dB] for $C_0=C(10)$ and
demonstrates the immense difference in the required powers. This
illustrates the fact that, for the same required secondary rate,
the interference towards the surrounding systems (both primary and
secondary) is markedly decreased when OIC is used.

\begin{figure}
  \centering
  \includegraphics[width=9cm]{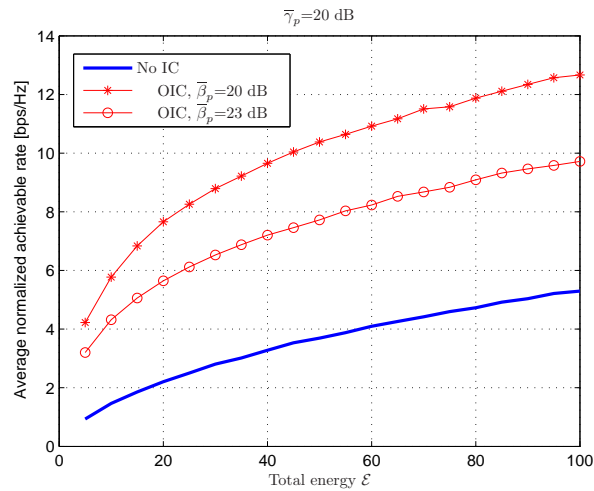}\\
  \caption{Normalized average achievable rate in [bps/Hz] as a
  function of the average SNR $\overline{\gamma}_s=\frac{{\cal E}}{\nu}$ on the secondary link. The number of channels in
  the system is $M=10$.
  }\label{fig:AverageRate1}
\end{figure}

\begin{figure}
  \centering
  \includegraphics[width=9cm]{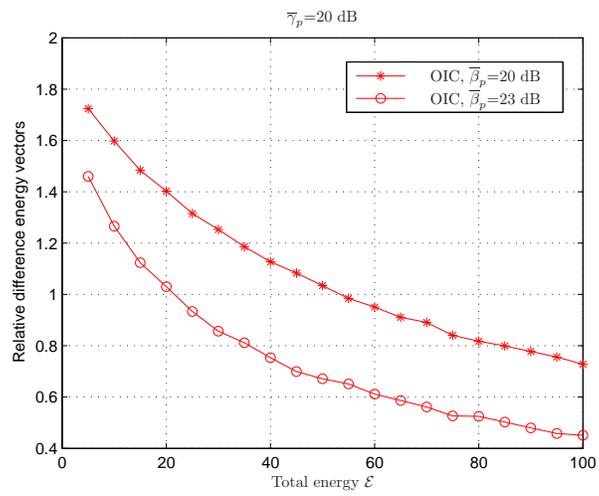}\\
  \caption{The average relative difference between the energy
  allocation vectors, calculated as $\frac{\sqrt{\|\textbf{E}_{oic}-\textbf{E}\|}}{{\cal
  E}}$.
  }\label{fig:Ediff}
\end{figure}

Figures~\ref{fig:AverageRate1} and~\ref{fig:Ediff} show the
evaluation results for a system with $M=10$ channels. When OIC is
used, the intercepted water--filling is applied. For the case
without interference cancellation, the conventional water--filling
is used. The abscissa depicts the scalar value of the total
applied energy by the secondary system. For a given value of
${\cal E}$, the normalized achievable rate is the sum of the rates
for all 10 channels (achieved with intercepted water--filling) and
the value is obtained by averaging over $10^4$ iterations. In each
iteration, the value $\nu_m$ for a given channel is generated as
$\nu_m=\frac{1}{\gamma_m}$, where $\gamma_m$ is exponentially
distributed variable with average value 1. This helps us to
interpret the energy in terms of SNR: The average secondary SNR
per channel is $\frac{{\cal E}}{M}$. Also, in each iteration, the
value $\gamma_{p,m}$ is generated randomly from an exponential
variable with mean value $\overline{\gamma}_p=20 dB$. The value
$\beta_{p,m}$ is generated randomly from an exponential variable
with mean value 20 dB and 23 dB, respectively, for each of the two
OIC curves. We can see that the opportunistic rate adaptation with
intercepted water--filling can lead to significant rate
improvements. As expected, when
$\overline{\beta}_p>\overline{\gamma}_p$ the secondary has less
opportunity to decode the primary signal, such that the
improvement over the case without interference cancellation is
decreased.
%\begin{figure}
%  % Requires \usepackage{graphicx}
%  \centering
%  \includegraphics[width=9cm]{_AverageRate1.eps}\\
%  \caption{Normalized average achievable rate in [bps/Hz] as a
%  function of the average SNR $\overline{\gamma}_s=\frac{{\cal E}}{\nu}$ on the secondary link. The number of channels in
%  the system is $M=10$.
%  }\label{fig:AverageRate1}
%\end{figure}

Fig.~\ref{fig:Ediff} reveals what is the difference in the energy
allocation between the conventional water--filling and the
intercepted water--filling. Let $\textbf{E}_{oic}$ denote the
energy allocation vector with intercepted water--filling, while
$\textbf{E}$ denotes the energy allocation vector for conventional
water--filling in the case no IC is applied. Clearly, for given
${\cal E}$, the sum of the components of each vector is equal to
${\cal E}$. The relative difference is calculated as
$\frac{\sqrt{\|\textbf{E}_{oic}-\textbf{E}\|}}{{\cal E}}$. When
the average required minimal SNR $\overline{\beta}_p=23 dB >
\overline{\gamma}_p$, the energy allocation vectors obtained with
the OIC are closer to the ones obtained without IC. This is
because, for higher $\overline{\beta}_p$, there is less chance
that a given channel will apply an intercepted water--filling. The
relative difference decreases as the energy increases. As stated
in the previous section, when the total energy is sufficiently
high, then the intercepted water--filling and the conventional
water--filling yield to identical energy allocation vectors, but
still different rates.
%\begin{figure}
%  % Requires \usepackage{graphicx}
%  \centering
%  \includegraphics[width=9cm]{_Ediff.eps}\\
%  \caption{The average relative difference between the energy
%  allocation vectors, calculated as $\frac{\sqrt{\|\textbf{E}_{oic}-\textbf{E}\|}}{{\cal
%  E}}$.
%  }\label{fig:Ediff}
%\end{figure}

\section{Conclusion}

We have investigated the problem of spectrally efficient operation
in a cognitive radio system under interference from a primary
system. A secondary (cognitive) receiver (RX) observes a multiple
access channel of two users, one user being the desired secondary
transmitter (TX) and the other the undesired primary TX. However,
the primary system selects the transmission rate independently of
the secondary system. If the link from the primary TX to the
secondary RX is weak, then the secondary RX is not able to decode
the primary signal. In order to make the best use of the power
over the secondary link, the secondary system should apply
\emph{Opportunistic Interference Cancellation (OIC)}. With OIC,
the secondary system cancels the interference from the primary
system whenever such opportunity is created by (a) selection of
the data rate in the primary system and (b) the link quality
between the primary TX and the secondary RX. We derive the
achievable data rate in the secondary system, which is a function
of the power applied by the secondary system. The parameters of
this function are the power received from the primary and the rate
applied in the primary system. We have also devised a method that
does not use time--sharing in order to achieve all the achievable
rate pairs in the multiple--access channel. This method has a
practical significance for a cognitive radio system, since it
enables rate adaptation without requiring the primary system to
perform a particular action. The derived rate adaptation function
is then applied in the scenario when the secondary system uses
multiple channels interfered by the primary. In this case, based
on the observed state in each of the available channels, the
secondary system should allocate the transmission power in a way
that maximizes the achieved sum--rate. Due to the features of the
derived rate adaptation function, the conventional water--filling
cannot be used. Therefore, we have introduced the method of
intercepted water--filling. We have presented numerical results
that illustrate the benefit of the devised methods of OIC and the
intercepted water--filling.

As a future work, we will consider the strategies for power/rate
adaptation when there are multiple concurrent cognitive radio
systems. Regarding the devised method of rate adaptation, we are
planning to quantify the improvement that it brings when we
consider finite packet length and practical (suboptimal)
modulation and coding methods.

%\bibliographystyle{IEEEtran}
%\bibliography{IEEEabrv,Dyspan_refs}
%----------------------------------------------------------------------

%----------------------------------------------------------------------

\end{document}